\documentclass[aps,pra,showpacs,superscriptaddress,twocolumn]{revtex4-1}
\usepackage{graphicx} 
\usepackage{amsmath}
 
\newcommand{\beq}{\begin{equation}}
\newcommand{\enq}{\end{equation}}
\newcommand{\bea}{\begin{eqnarray}}
\newcommand{\ena}{\end{eqnarray}}
\newcommand{\la}{\langle}
\newcommand{\ra}{\rangle}

\begin{document}
\title{Mott insulators in plaquettes}
\author{J.-P. Martikainen}
\email{Jani-Petri.Martikainen@aalto.fi}
\affiliation{COMP Centre of Excellence, Department of Applied Physics, Aalto University, Fi-00076 Aalto, Finland}
%\author{P. T\"{o}rm\"{a}}
%\affiliation{Aalto University, P.O. Box 1510, Fi-00076 Aalto, Finland}
\date{\today}

\begin{abstract}
We study small systems of strongly interacting ultracold atoms under 
the influence
of gauge potentials and spin-orbit couplings. We use second order
perturbation theory in tunneling, derive an effective theory 
for the strongly correlated insulating states
with one atom per site, and solve it exactly.
We find dramatic changes in the level structure and in the amount of degeneracies expected. We
also demonstrate the dynamical
behavior as the barriers between plaquettes are gradually removed
and find potentially high overlap with 
the resonating valence bond (RVB) state of the larger system.
\end{abstract}

\maketitle 
\section{Introduction}
\label{sec:introduction}
Weakly interacting gas of bosons can form a Bose-Einstein 
condensate at low enough temperatures.
This condensate can be often well described by a classical theory where condensate
atoms are described by a condensate wavefuction whose dynamics follows from the Gross-Pitaevskii
equation~\cite{Dalfovo1999a}. 
In optical lattices potential barriers between sites suppress the movement
of atoms and increase confinement which increases effective interaction strengths. Consequently
optical lattices can push the weakly interacting quantum system into the regime a strongly correlated
physics. Bose-Einstein condensation of weakly interacting systems gives way to Mott-insulator
physics at high interactions and this transition was experimentally observed
by Greiner {\it et al.}~\cite{Greiner2002a}.

While the position of the transition (in higher dimensional systems)
can be fairly accurately predicted with the Gutzwiller ansatz that is a product
state of individual site quantum states, it predicts the physics of Mott insulators
poorly since it does not take into account higher order tunneling processes. However
these strongly correlated insulators can be studied in terms of different spin models and
can have a very rich physics of quantum magnetism and criticality~\cite{Sachdev2008a}.
At a formal level spin models can be derived from the second order perturbation theory in
kinetic (tunneling) energy of the lattice model~\cite{Kuklov2003a,Altman2003a}.
 
Recently there has been rapid experimental progress in creating artificial 
spin-orbit couplings~\cite{Lin2011a}
as well as gauge potentials~\cite{Struck2011a,Struck2012a} 
in ultra cold quantum gases. 
Given this background even non-Abelian gauge fields~\cite{Ruseckas2005a,Hauke2012a}
have become feasible. (For a recent review see, for example, Dalibard {\it et al.}~\cite{Dalibard2011a}.)
In addition to these advances, a three-dimensional array of independent plaquettes
has also been demonstrated even to a level of being able to prepare
specific quantum states in each plaquette~\cite{Nascimbene2012}.

The purpose of this article is to study insulating states in an optical lattice
for small systems by exactly diagonalizing the effective Hamiltonian corresponding
to the second order perturbation theory in tunneling. We will solve the system under quite
general conditions so that tunneling coefficients can be spatially varying and
artificial gauge potentials included easily. We demonstrate the level structure
for few simple choices of phases and point out when, for example, the amount of degeneracies
can vary with gauge potentials.
Furthermore, we show some examples of dynamics in multi-plaquette systems and
solve the problem of plaquettes with spin-orbit interactions exactly
for $2$- and $4$-sites.

There are some interesting recent
studies on the phase diagrams of 
Mott insulators with spin-orbit couplings~\cite{Cole2012a,Radic2012a} that pointed out
how the effective Hamiltonian becomes
a combination of Heisenberg model, quantum compass model, and Dzyaloshinskii-Moriya
interactions. The approach used in this paper is different in that our focus is
on the full solution for small systems which naturally enable also the solution of time-dependent
problems. This is relevant since in the experiment by 
Nascimb\'{e}ne {\it et al.}~\cite{Nascimbene2012}
resonating valence bond states were prepared in plaquettes of an 
optical lattice. In these experiments dynamics was an important ingredient and since the system
was made out of independent plaquettes, exact diagonalization for small systems provides
a natural theoretical framework.

The paper is organized as follows. We begin by outlining the 
theory relevant for our purposes in Sec.~\ref{sec:theory}. 
In Sec.~\ref{sec:gauge_potentials} we exactly diagonalize the system with some simple gauge
potentials both for the square as well as for the triangular plaquette.
Then in Sec.~\ref{sec:few_plaquettes} we explore the dynamical behavior when barriers between
plaquettes are removed and in
Sec.~\ref{sec:so_coupled} we exactly diagonalize the Mott insulating plaquette with spin-orbit
interactions.
 We end with a few concluding remarks in Sec.~\ref{sec:conclusions}.

\section{Formalism}
\label{sec:theory}
In each site ${\bf r}$  of the lattice
the local physics for bosons is described by the interaction Hamiltonian~\cite{Kuklov2003a,Altman2003a}
\begin{eqnarray}
H_{I,{\bf r}}&=&\frac{U_{11}}{2}\hat n_{1,{\bf r}}(\hat n_{1,{\bf r}}-1)
+\frac{U_{22}}{2}\hat n_{2,{\bf r}}(\hat n_{2,{\bf r}}-1)\nonumber \\
&+&U_{12}\hat n_{1,{\bf r}}\hat n_{1,{\bf r}}, %\nonumber
\end{eqnarray}
where $\hat n_{\alpha,{\bf r}}=\hat\psi_{\alpha,{\bf r}}^\dagger\hat\psi_{\alpha,{\bf r}}$ ($\alpha\in\{1,2\}$) are the density
operators for each component. Here the onsite eigenstates can be 
written in terms
of occupation numbers as $|n_1,n_2\ra_{\bf r}$.
The approach works equally well for fermions, but then
$U_{11}=U_{22}=0$ since $s$-wave interaction
between identical fermions vanishes and Pauli exclusion principle must be enforced. Furthermore,
this problem can be straightforwardly generalized 
to spinorial bosonic systems, multi-component
fermionic systems~\cite{Krauser2012a}, 
or Bose-Fermi mixtures~\cite{Best2009a,Shin2008b}
by simply changing the term describing onsite interactions~\cite{comment2012a}.

In this work tunneling processes are kept very general so that they can
include position dependent tunnelings as well as, later in the text,
spin-orbit couplings. In the absense of spin-orbit couplings the
tunneling processes are described by
\beq
H_T=\sum_{\alpha,m}
-t_{\alpha,m}\hat\psi_{\alpha,m+}^\dagger\hat\psi_{\alpha,m-}
-t_{\alpha,m}^*\hat\psi_{\alpha,m-}^\dagger\hat\psi_{\alpha,m+},
\enq
where $\alpha$ indicates the component index, $m$ indicates the bond,
while $\hat\psi_{\alpha,m+}$ and $\hat\psi_{\alpha,m-}$ are the field operators
for atoms of type $\alpha$ on either end of the bond $m$. Later we will 
introduce also spin-orbit couplings in which case the atom type 
can change in the tunneling process.

Throughout this paper we focus on the 
experimentally simplest case of having one atom per site. For all states having just
one atom per site interaction energy vanishes so double (and higher) occupancy
is energetically unfavorable. Adding tunneling terms into the model implies
removing an atom from one site and adding it into another. If the initial
state only had singly occupied sites,
this implies that tunneling process creates a state where one site is
doubly occupied. In the lowest order perturbation theory the tunneling 
term does not give rise to any energy shift since the state with double occupancy
is orthogonal to initial state, but at second order there can be another
tunneling process which brings the state back into the initial subspace of singly
occupied sites. At this level tunneling terms do matter and they determine the energetics
and dynamics of the various possible Mott insulating states.

When the tunneling processes are described by a Hamiltonian $H_T$ this
second order perturbation theory gives rise to an effective Hamiltonian~\cite{Radic2012a}
\beq
(H_{{\rm eff}})_{\alpha\beta}=-\sum_{\gamma}\frac{(H_T)_{\alpha\gamma}(H_T)_{\gamma\beta}}
{E_\gamma-(E_\alpha+E_\beta)/2},
\enq
where $E_i$ are the eigenenergies of the interaction Hamiltonian, $\alpha$ and $\beta$ live
in the manifold with one atom per site, while states labeled by 
$\gamma$ have one site with $2$ atoms per site.

In the literature the this expression is typically written
in terms of pseudo-spin operators. While this can formally be done
in our cases as well, the expressions become ever more complex as
tunneling properties in the system become more diverse. For this reason
we choose not to rephrase the problem in terms of spin-operators,
but construct the relevant Hamiltonians 
in the subspace with one atom per site with the computer. In this way
all associated ``spin''-models are included, but as an input one only
needs the onsite interactions and tunneling coefficients. In principle,
both interactions as well as tunnelings could be position dependent. The latter
option is in fact important when we wish to study effects of gauge potentials
in lattices and position dependence of atom-atom interactions can be important,for example,
in bipartite lattices~\cite{Martikainen2012a}. 
If we were to write results in terms of pseudo-spins, we
would have to derive a separate spin-model for each variant 
of the problem.

Our procedure amounts to first constructing all eigenstates
of $H_I=\sum_{{\bf r}} H_{I,{\bf r}}$ with (on average) unit filling and at maximum
$2$ atoms per site. From these states we construct the subspace with exactly
one atom per site and then compute the elements $H_{eff}$ numerically. The resulting
Hamiltonian can then be diagonalized and this solution (if so desired) 
can then be used to solve dynamical problems as well. 
In this paper we will focus on $2\times 2$ or triangular plaquettes
since the first system has been experimentally realized~\cite{Nascimbene2012}
and the latter one can be realized~\cite{Eckardt2010a}. Furthermore,
both can be solved fully without difficulty. We also solve dynamical problems in
systems of two and three weakly coupled plaquettes.

\section{Mott insulators in gauge potentials}
\label{sec:gauge_potentials}
\begin{figure}
%\begin{tabular}{ll}
\includegraphics[width=0.49\columnwidth]{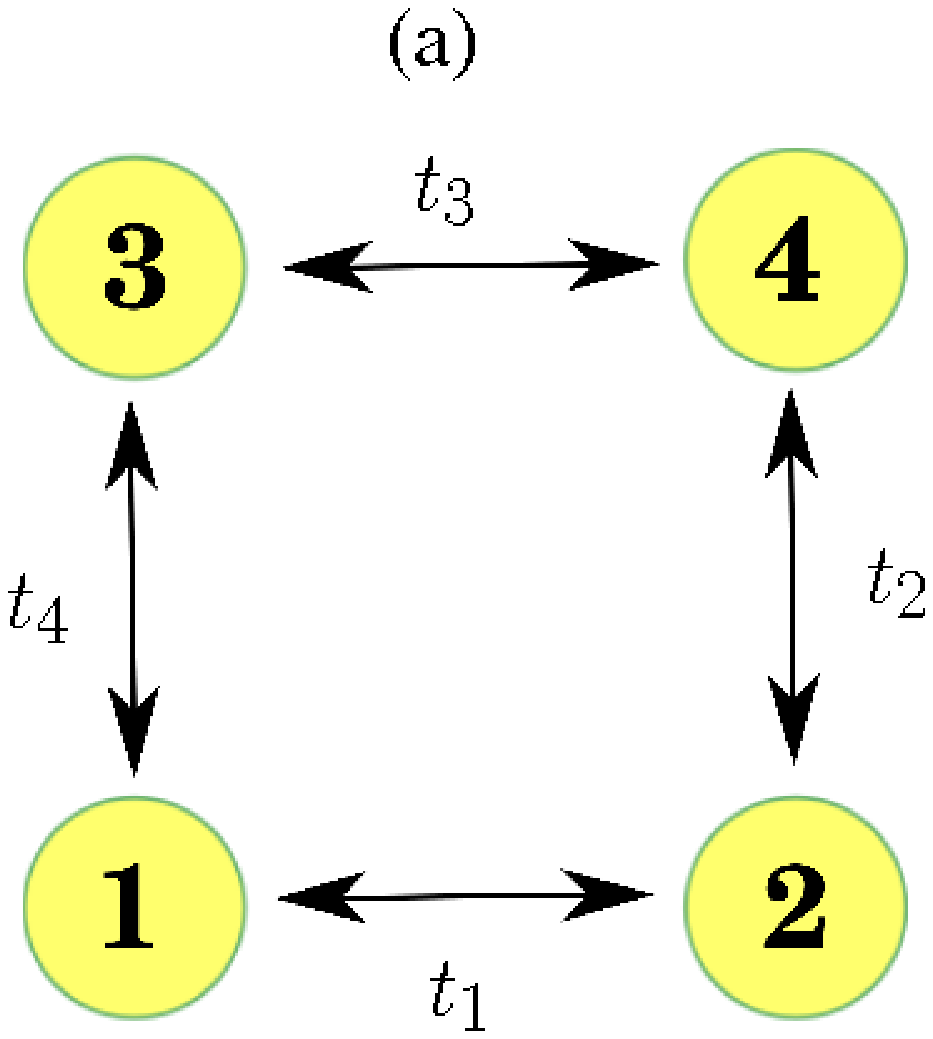}
\includegraphics[width=0.42\columnwidth]{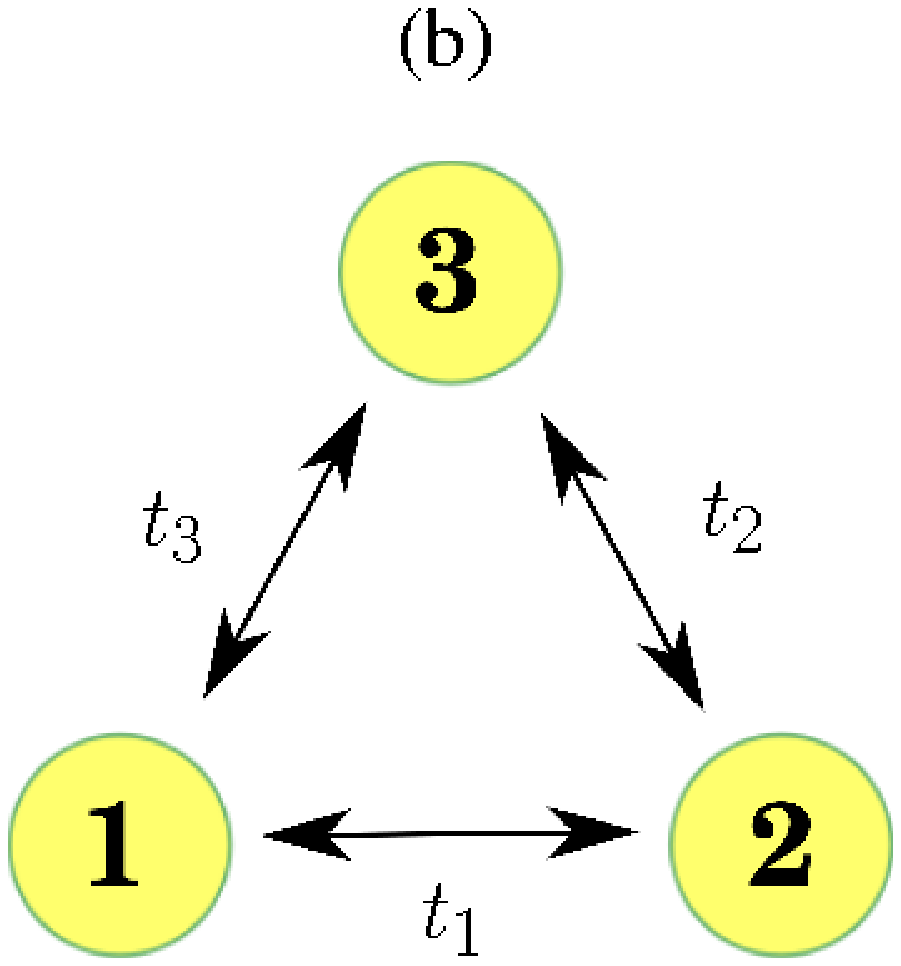}
%\end{tabular}
\caption[Fig1]{(a) Schematic description of the $2\times 2$ plaquette
and notation used in the text. (b) Schematic description a triangular plaquette.
}
\label{fig:schematic}
\end{figure}
We show a schematic description of our system(s) elementary cells in Fig.~\ref{fig:schematic}.
The various tunneling coefficients are given by $t_m$ and they describe processes
like (example for $t_1$)
$-t_{\alpha,1}\hat\psi_{\alpha,2}^\dagger\hat\psi_{\alpha,1}-t_{\alpha,1}^*\hat\psi_{\alpha,1}^\dagger\hat\psi_{\alpha,2}$.
for tunneling of $\alpha$ atoms between sites $1$ and $2$. Nascimb\'{e}ne 
{\it et al.}~\cite{Nascimbene2012}
solved the eigenstates of this system in the case when $t_2=t_4=t_y$ is different from $t_1=t_3=t_x$.
We wish to understand the role of gauge potentials in such a system and for this reason
we allow for the possibility of complex and position dependent tunneling coefficients.
In contrast to superfluid regime, 
if both components experience the same gauge potential, gauge potential does not affect the 
Mott insulator. For gauge potentials to play a role, we have to allow for the possibility of
spin dependent gauge potential. To keep things as simple as possible we choose
$t_{1,1}=e^{i\phi}/2$, all other tunneling coefficients as $1/2$, and vary the phase
$\phi$. (Using Peirls substitution the phase could be related to 
gauge vector potential ${\bf A}$ through
$\phi=\int_{{\bf r}_i}^{{\bf r}_j} d{\bf r}\cdot {\bf A}/\hbar$).
This gives rise to the Harper
model  in a plaquette~\cite{Gerbier2010a} and if the 
flux $\alpha=\phi/2\pi=p/q$ 
is rational the ideal theory has a $q$-fold degenerate ground state.

\begin{figure}
%\begin{tabular}{ll}
\includegraphics[width=0.49\columnwidth]{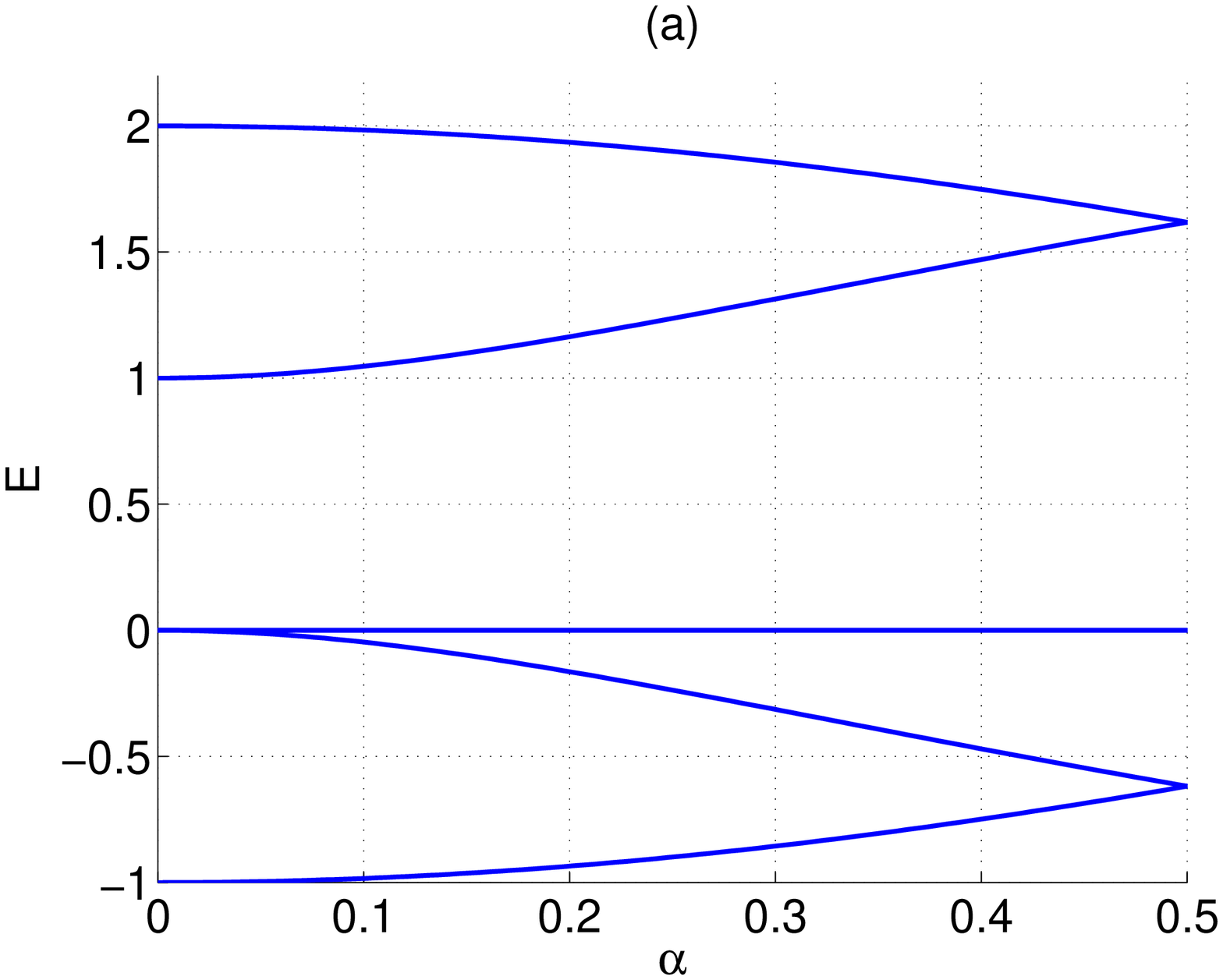}
\includegraphics[width=0.48\columnwidth]{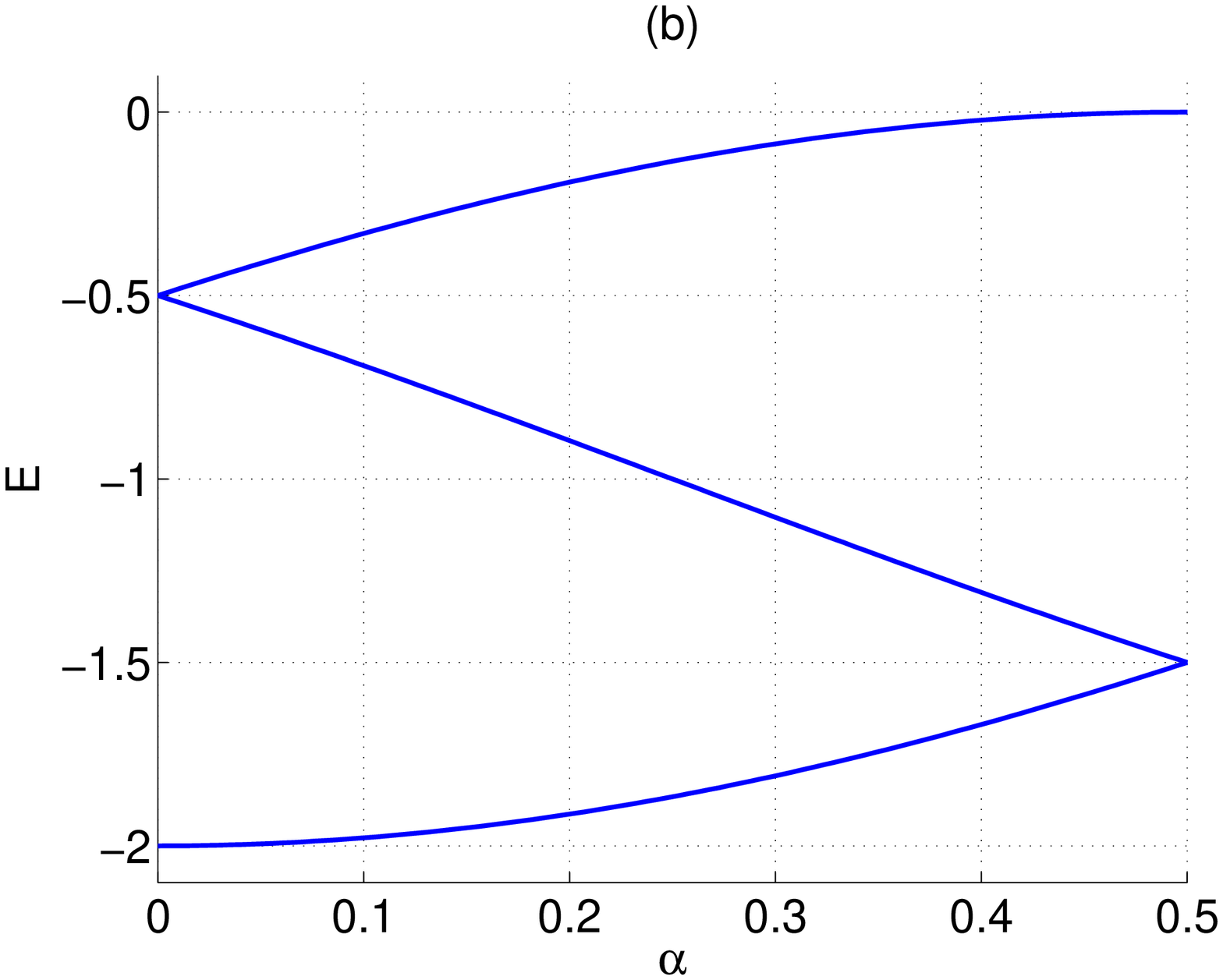}
%\end{tabular}
\caption[Fig2]{(a) Eigenenergies of $2\times 2$ plaquette when another component 
experiences a gauge potential and $N_1=N_2$.
The uppermost state at $\alpha=0$ is the s-wave RVB state and the first excited state
is the d-wave RVB state as discussed in Ref.~\cite{Nascimbene2012}.
We choose $|t_i|=1/2$ and $U_{11}=U_{22}=U_{12}=1$.
(b)  Eigenenergies in a triangular plaquette when another component 
experiences a gauge potential and $N_1=N_2+1$. 
}
\label{fig:gaugepot}
\end{figure}

In Fig.~\ref{fig:gaugepot} (a) we show the eigenenergies as a function of $\alpha$. 
At $\alpha=0$ the result agrees with those reported in Ref.~\cite{Nascimbene2012}, but
as gauge potential is turned on the $3$-fold degeneracy of the fist excited state
is broken as the $d$-wave RVB state splits into a separate branch. 
As gauge potential becomes stronger the ground state becomes doubly
degenerate at $\alpha=1/2$ also in the interacting system. Furthermore, the 
gap separating the highest $s$-wave RVB state closes at $\alpha=1/2$ and the eigenstate 
there becomes doubly degenerate.

One way to create gauge potentials in lattices is to shake 
the lattice~\cite{Eckardt2010a,Sacha2012a} appropriately
and such tools have indeed been demonstrated 
experimentally~\cite{Lignier2007a,Chen2011a,Ma2011a,Struck2012a}
. Some of the more
simple methods of shaking cannot create a magnetic flux in a square lattice while
they might do so in a triangular lattice. For this reason it is instructive to
solve the problem also for a triangular plaquette. We show a typical the result
in  Fig.~\ref{fig:gaugepot} (b). In a triangular plaquette with one atom per site
the different component have necessarily different atom numbers. If $N_1=N_2+1$ there
are three eigensolutions. The excited state is doubly degenerate in the absence of 
gauge potential while its degeneracy is broken in the gauge potential. At $\alpha=1/2$
the ground state is again doubly degenerate while the excited state is non-degenerate.

%At $\phi=\pi$ and  $\phi=\pi/2$ ($\alpha=1/2$ and $\alpha=1/4$) 
%we get the same dispersions, by making just $t_1=te^{i\phi}$ as by
%having ${\bf t}=t(1,e^{\phi/2},e^{2\phi/2},e^{3\phi/2})$. However, generally these do not agree.

\section{Few coupled plaquettes}
\label{sec:few_plaquettes}
In the experiment by Nascimb\'{e}ne {\it et al.}~\cite{Nascimbene2012} the
plaquettes were not strongly coupled and they were able to create the
$s$-wave RVB state in each plaquette. We can use our approach  
to also study systems with more than one plaquette. In this case
it becomes possible to study how coupling between plaquette changes the system
behavior. Since we have not made any assumptions about the magnitudes of tunneling strengths
we can simply add tunneling terms between plaquettes without, for example, a need for
a new perturbation theory in ``small'' inter-plaquette tunneling.
One notable result from solving the problem for coupled plaquettes 
(for bosons) is that the ground state
is degenerate in the limit of vanishing inter-plaquette tunneling
and is separated from the first excited states by only a small gap as inter-plaquette
tunneling is introduced. 

In sharp contrast to this, the most excited state corresponding
to the product state of plaquette $s$-wave RVB states is gapped in 
the limit of vanishing inter-plaquette tunneling
and the gap remains quite large even as the inter-plaquette tunneling approaches the 
in-plaquette tunneling. More quantitatively, when  all coupling strengths are the 
same, $U_{11}=U_{22}=U_{12}=U$, and
in-plaquette tunneling is taken as $t=1/2$ the 
gap $\Delta E=1/U$ when inter-plaquette tunneling is zero. When inter-plaquette tunneling
equals in-plaquette tunneling the gap is reduced to $\Delta E\approx 0.77/U$ for two plaquettes
and $\Delta E\approx 0.65/U$ for three. This separation of the (experimentally relevant) most excited
state might simplify studies where Hamiltonian is changed as a function of time, since
unwanted transitions to other states can be suppressed. 

\begin{figure}
%\begin{tabular}{ll}
\includegraphics[width=0.49\columnwidth]{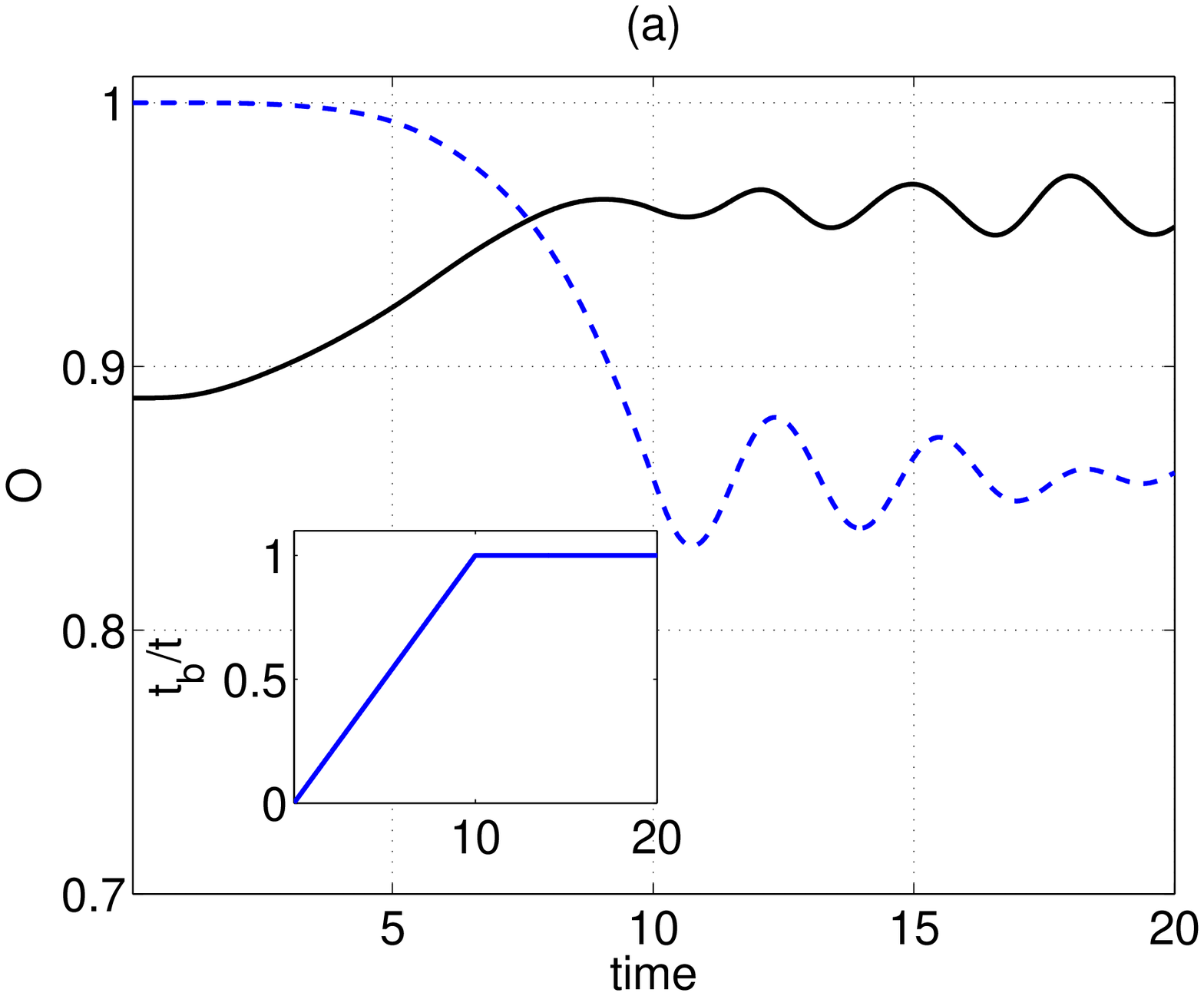}
\includegraphics[width=0.48\columnwidth]{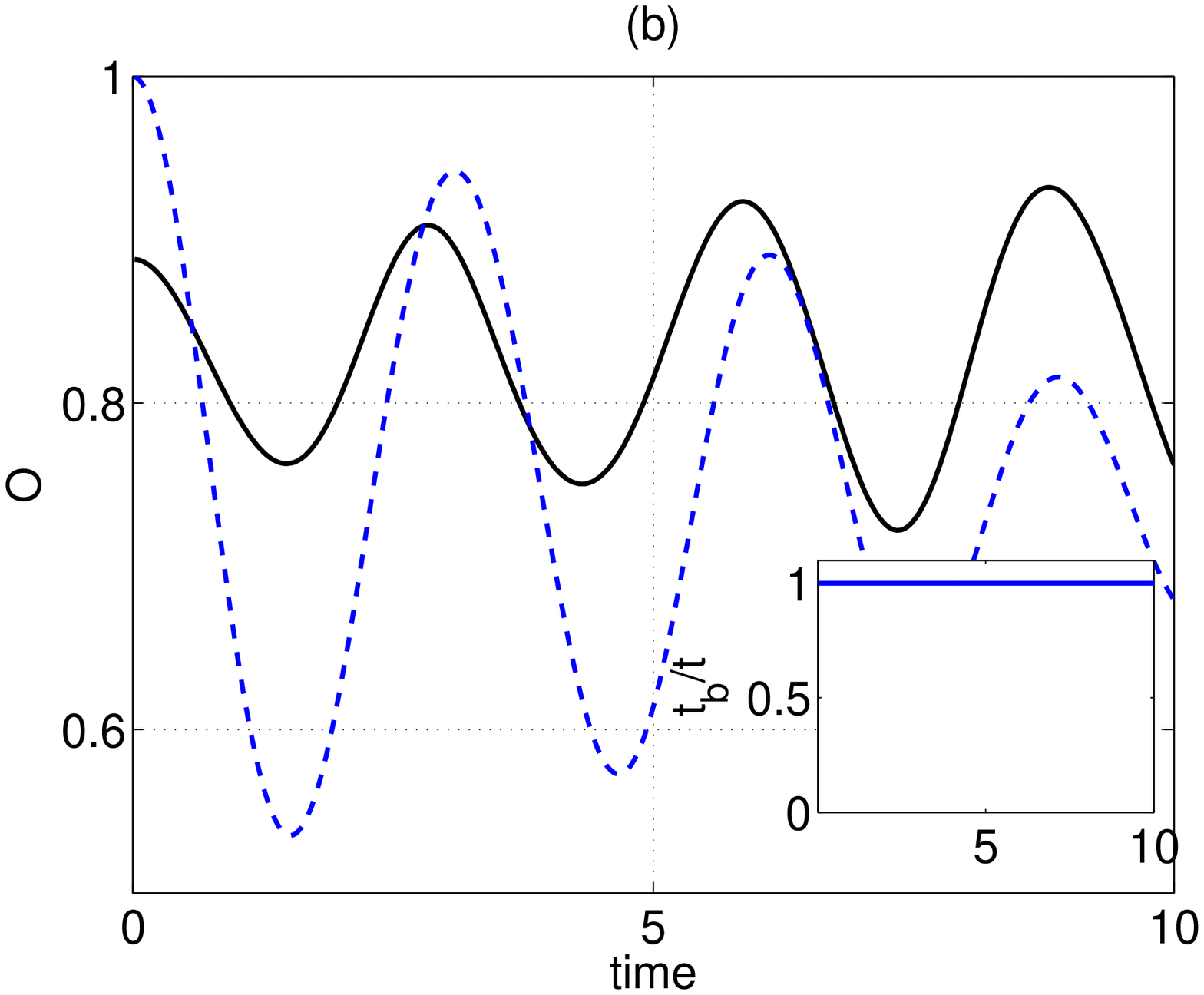}
%\end{tabular}
\caption[Fig3]{Overlap $O=|\la RVB|\psi(t)\ra|^2$ with the RVB-state over all $8$-sites (solid line) and
with the initial state $|\psi_0\ra$
that was a product state of individual plaquette $s$-wave RVB-states
(dashed line).
Insets shows the strength of the 
inter-plaquette tunneling $t_b$ relative to in-plaquette tunneling $t$. 
In (a) we choose to ramp up
the inter-plaquette tunneling strength 
linearly and the kept it steady for the remainder of the simulation. In (b) the inter-plaquette
tunneling was abruptly turned on to be equal to in-plaquette tunneling.
We used $U_{11}=U_{22}=U_{12}=1$ and $t=1/2$, while the  unit of time was $\tau=1/U_{11}$.
}
\label{fig:2plaquette dynamics}
\end{figure}

To understand coupled plaquettes better, let us then demonstrate the dynamical behavior for
two coupled plaquettes. %Size of the Hilbert space=70 
We take the initial state to be the most excited state for uncoupled
plaquettes. This state corresponds to the product state of plaquette $s$-wave RVB states.
We then either increase the inter-plaquette tunneling $t_b$
by linearly ramping it to the same
value as the in-plaquette tunneling $t$ or by turning it on instantaneously.
In Fig.~\ref{fig:2plaquette dynamics} we show how overlaps with the initial state
and the RVB-state over all $8$ sites (equal superposition of all possible ways to cover the
lattice with singlet bonds between nearest neighbors) behave. 
As can be seen, the initial state has a substantial overlap with the
resonating valence bond state in the $8$-site system and this overlap increases further
as  the inter-plaquette tunneling approaches the in-plaquette tunneling and bonds
start forming also between plaquettes. In this example, the 
maximum overlap is around $97\%$. The overlap with the initial product state
remains large, but is substantially smaller than the overlap with the RVB-state. 
Incidentally the overlap of the RVB-state with the most excited state of the $8$-site system
when inter-plaquette tunneling 
is equal to in-plaquette tunneling is around $96\%$. Our result with the linear ramp demonstrates
that the most excited state of the uncoupled plaquettes evolves adiabatically very
close to the most excited state of two strongly coupled plaquettes.

In Fig.~\ref{fig:2plaquette dynamics} (b) we show that if the barriers
between plaquettes are removed suddenly the system, unexpectedly, strays further away
from the initial state and dynamical behavior is more dramatic.
These results 
demonstrate that at least for small systems RVB-states can be formed with fairly high fidelity
starting from a collection of RVB-states in independent plaquettes and lowering the barriers
between plaquettes.

\section{Plaquettes with spin-orbit coupling}
\label{sec:so_coupled}
Recently it has become possible to create spin-orbit couplings in 
ultracold atomic gases~\cite{Lin2011a}. This exciting development
has opened up entirely new possibilities and motivated a rapidly expanding
literature on various aspects of spin-orbit coupled systems using cold gases.
The presence of 
spin-orbit coupling (SOC) changes our previous discussion in that 
it becomes possible for atoms to change type as they tunnel between sites.
The tight binding model for a spin-orbit coupled two-component 
gas is given by ($i,j\in \{1,2\}$)
\beq
H_T=-\sum_{{\bf r}ij}\sum_{\gamma=x,y}\left[
T_{{\bf r}\gamma}^{(i,j)}\hat\psi_{{\bf r},i}^\dagger \hat\psi_{{\bf r}+{\bf \eta}_\gamma,j}
+H.c.\right],
\enq
where $\hat\psi_{{\bf r},i}^\dagger$ creates an atom of type $i$ in the lattice site
${\bf r}$ and ${\bf \eta}_\gamma=a\hat\gamma$ is the vector to the nearest
neighbor of ${\bf r}$ along the direction $\gamma$ ($a$ is lattice spacing). 
$T_{{\bf r}\gamma}^{(i,j)}$ are the tunneling
matrices, but note that (as in previous section) 
since tunneling coefficients in the plaquette can be made to vary,
we have kept the possibility of position dependence in them.
In our notation we mostly follow Radic {\it et al.}~\cite{Radic2012a}.

The free spin-orbit coupled system can be expressed in terms of a 
vector potential ${\bf A}$ with a Hamiltonian $H_0=({\bf p}-{\bf A})^2/2m$, where
$m$ is the atomic mass. We choose the vector potential to be position independent
and given by
\beq
{\bf A}=\left(-m\alpha\hat\sigma_x, -m\beta\hat\sigma_y\right),
\enq
where $\hat\sigma_\gamma$ ($\gamma\in \{x,y,z\}$) are the Pauli matrices.
Peirls substitution
\beq
T_\gamma=t_\gamma e^{-iaA_\gamma}=t_\gamma e^{-i\theta_\gamma \hat\sigma_\gamma}
\enq
can then be used to relate tunneling coefficient $t_\gamma$ in
the absence of SOC (i.e. $\theta_x=\theta_y=0$)
to those with SOC. This substitution is not exact and can be
improved, but it is reasonably accurate over wide range of parameters and where
its accuracy is somewhat worse it is still qualitatively useful~\cite{Radic2012a}.
Inaccuracies can become more serious in and closer to the superfluid regime.

When the expressions for tunneling elements in SOC system are expanded we find
\beq
T_x=t_x\left(\begin{array}{cc}
\cos\theta_x & i\sin\theta_x \\
i\sin\theta_x & \cos\theta_x
\end{array}\right)
\enq
and
\beq
T_y=t_y\left(\begin{array}{cc}
\cos\theta_y  & \sin\theta_y \\
-\sin\theta_y & \cos\theta_y
\end{array}\right).
\enq

\begin{figure}
\includegraphics[width=0.8\columnwidth]{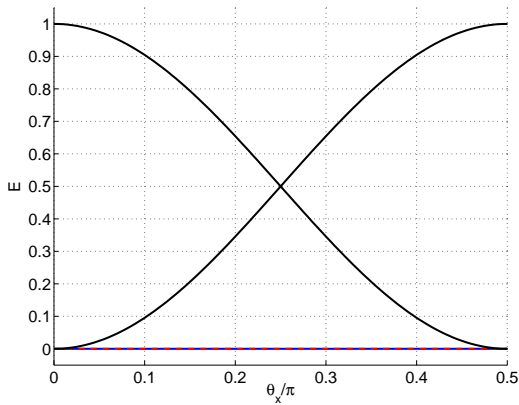}
\caption[Fig4]{(Color online) $2$-site system with spin-orbit coupling. 
The lowest level is doubly degenerate. We choose $t_x=1/2$ and $U_{11}=U_{22}=U_{12}=1$.
}
\label{fig:SO2site}
\end{figure}

In Fig.~\ref{fig:SO2site} we show the eigenenergies of the spin-orbit coupled
system with only $2$-sites. The basis is spanned by $4$ states and without spin-orbit coupling the 
ground state in $3$-fold degenerate. Spin-orbit coupling breaks this degeneracy so that
ground state remains $2$-fold degenerate. These ground states do not change with 
spin-orbit coupling, but the $2$ excited state energies do depend in the strength of SO coupling.
The eigenstates are however always such that
the lower one is of type $|\psi\ra=(|10\ra_1|10\ra_2-|01\ra_1|01\ra_2)/\sqrt{2}$
while the upper one is of type
$|\psi\ra=(|10\ra_1|01\ra_2-|01\ra_1|10\ra_2)/\sqrt{2}$.

\begin{figure}
\includegraphics[width=0.45\columnwidth]{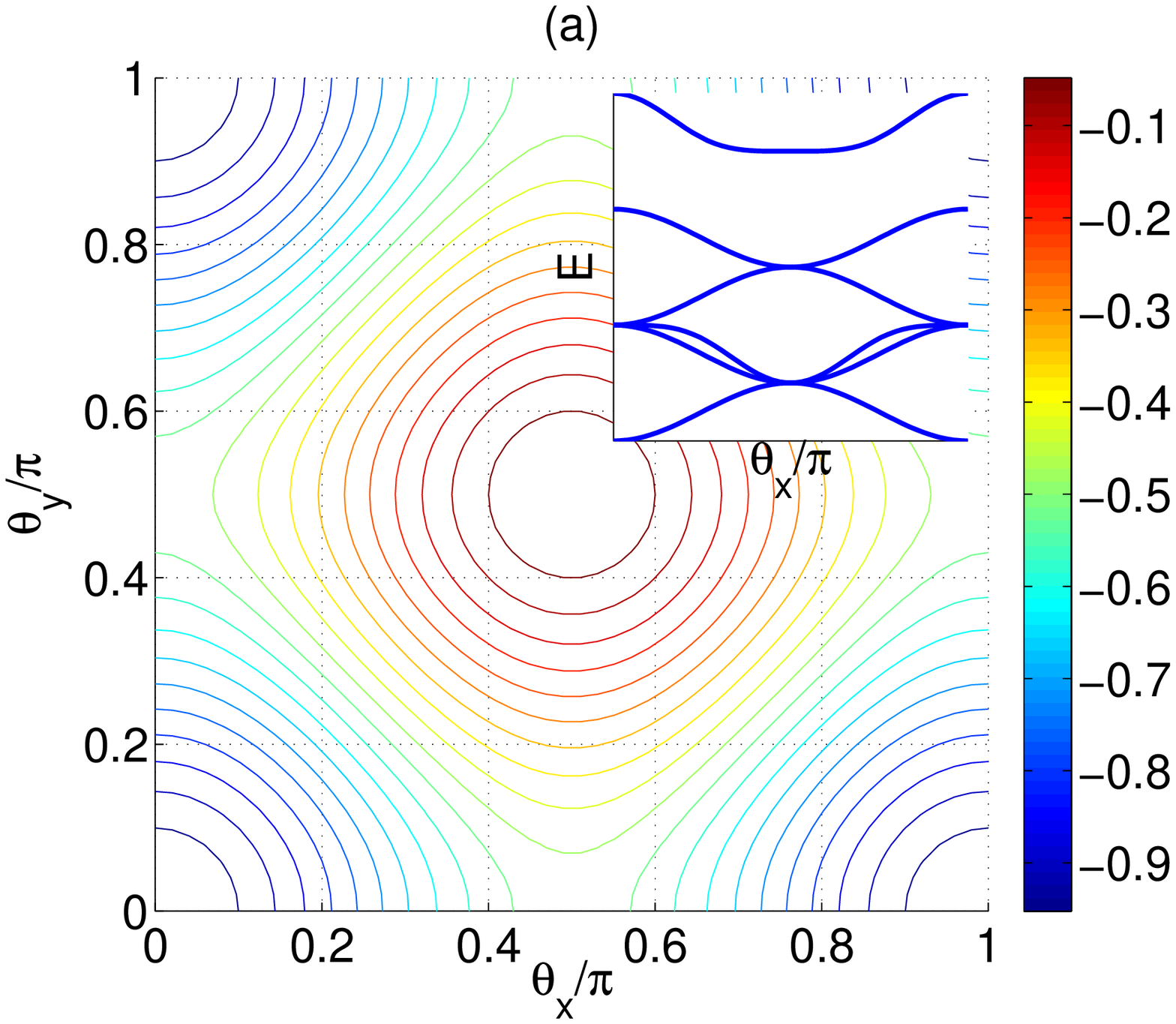}
\includegraphics[width=0.43\columnwidth]{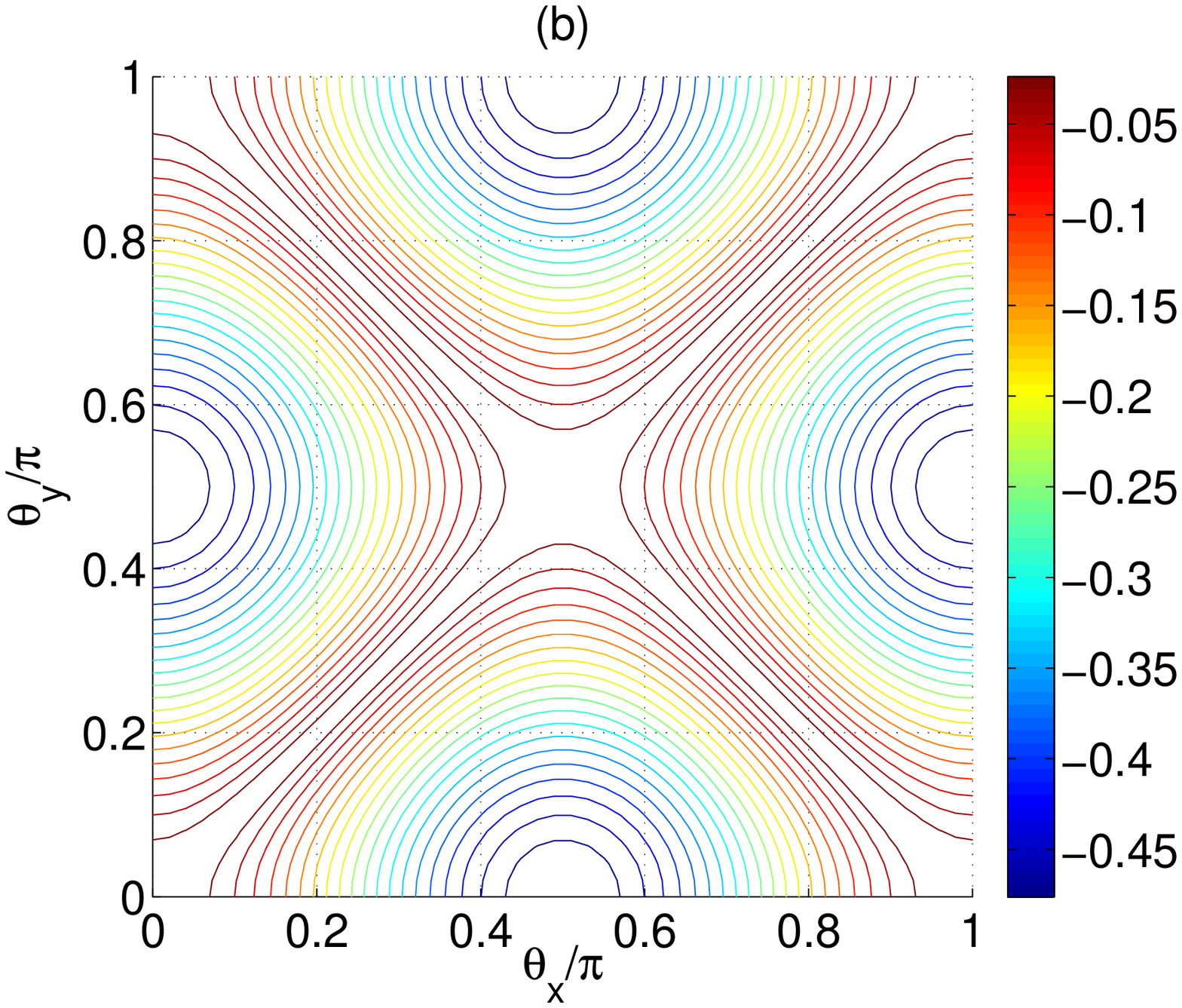}\\
\includegraphics[width=0.45\columnwidth]{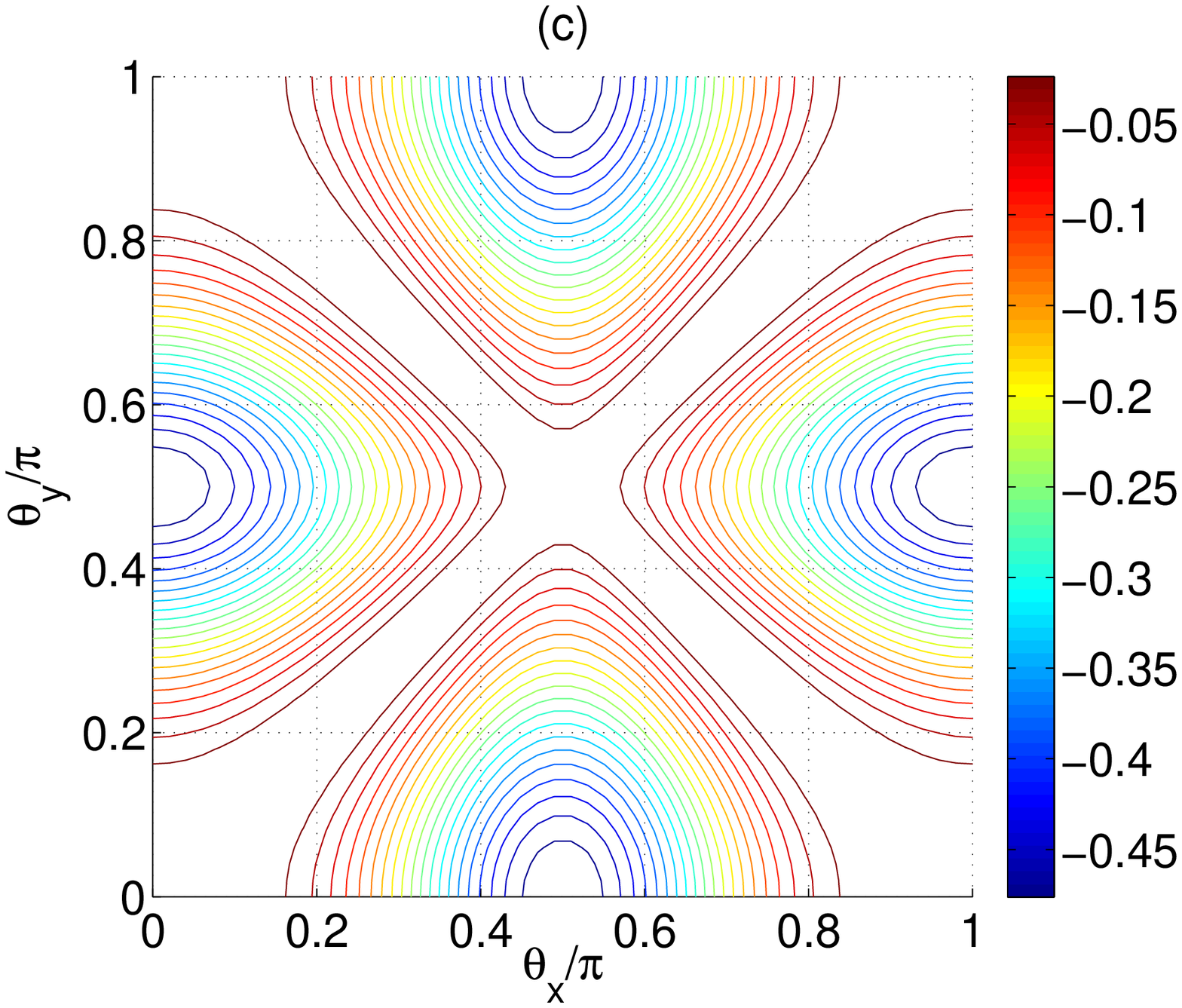}
\includegraphics[width=0.43\columnwidth]{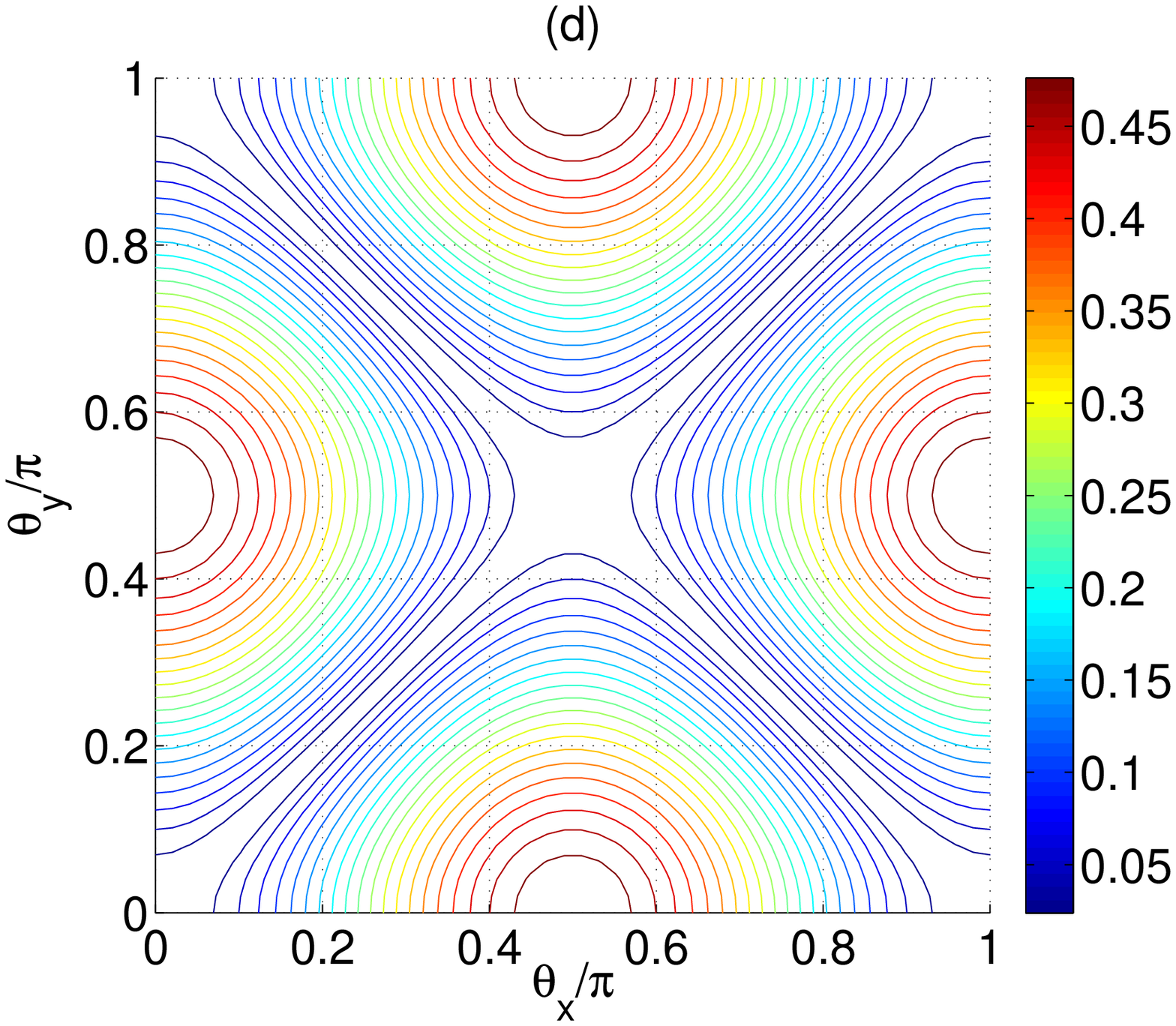}\\
\includegraphics[width=0.45\columnwidth]{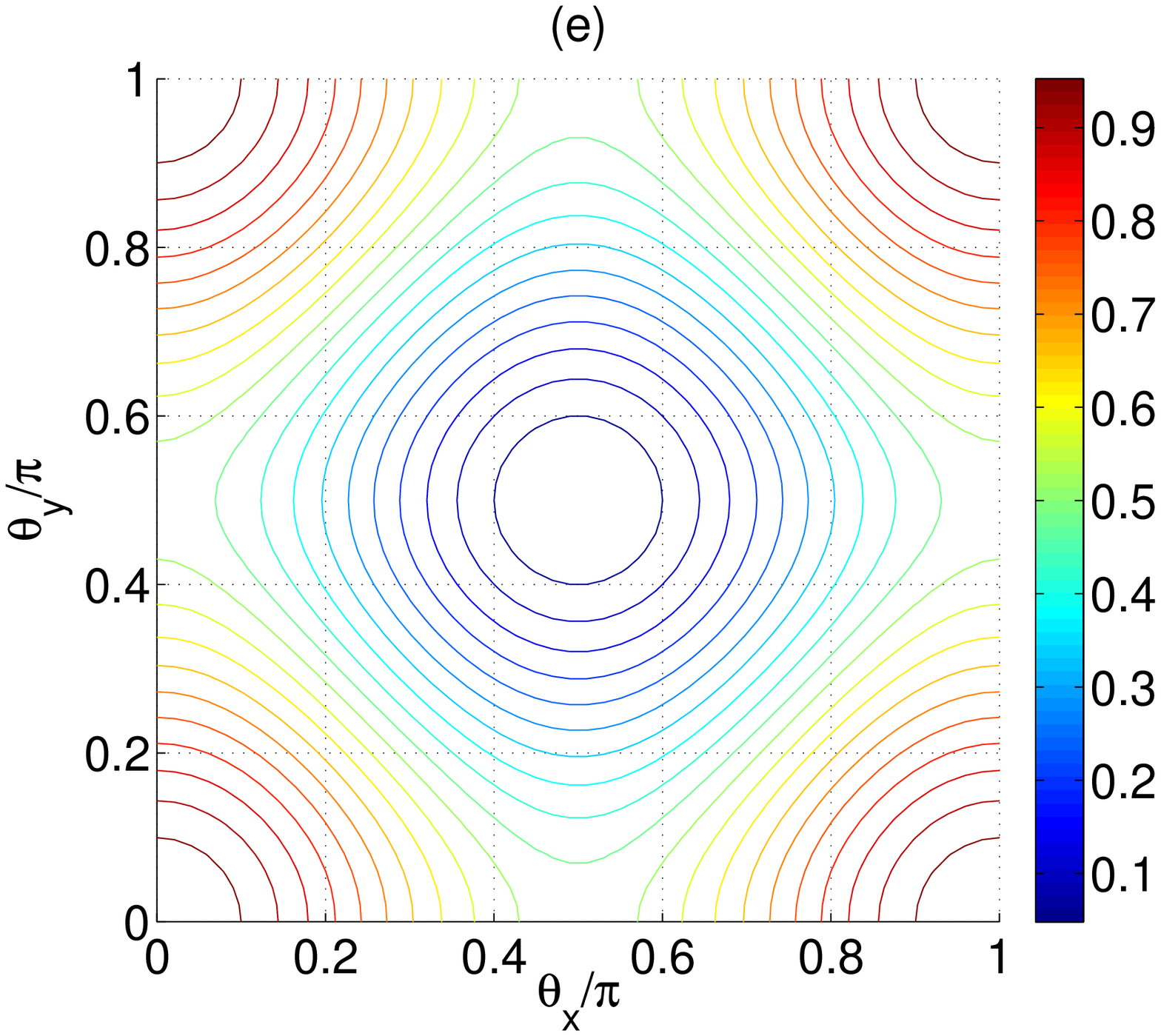}
\includegraphics[width=0.43\columnwidth]{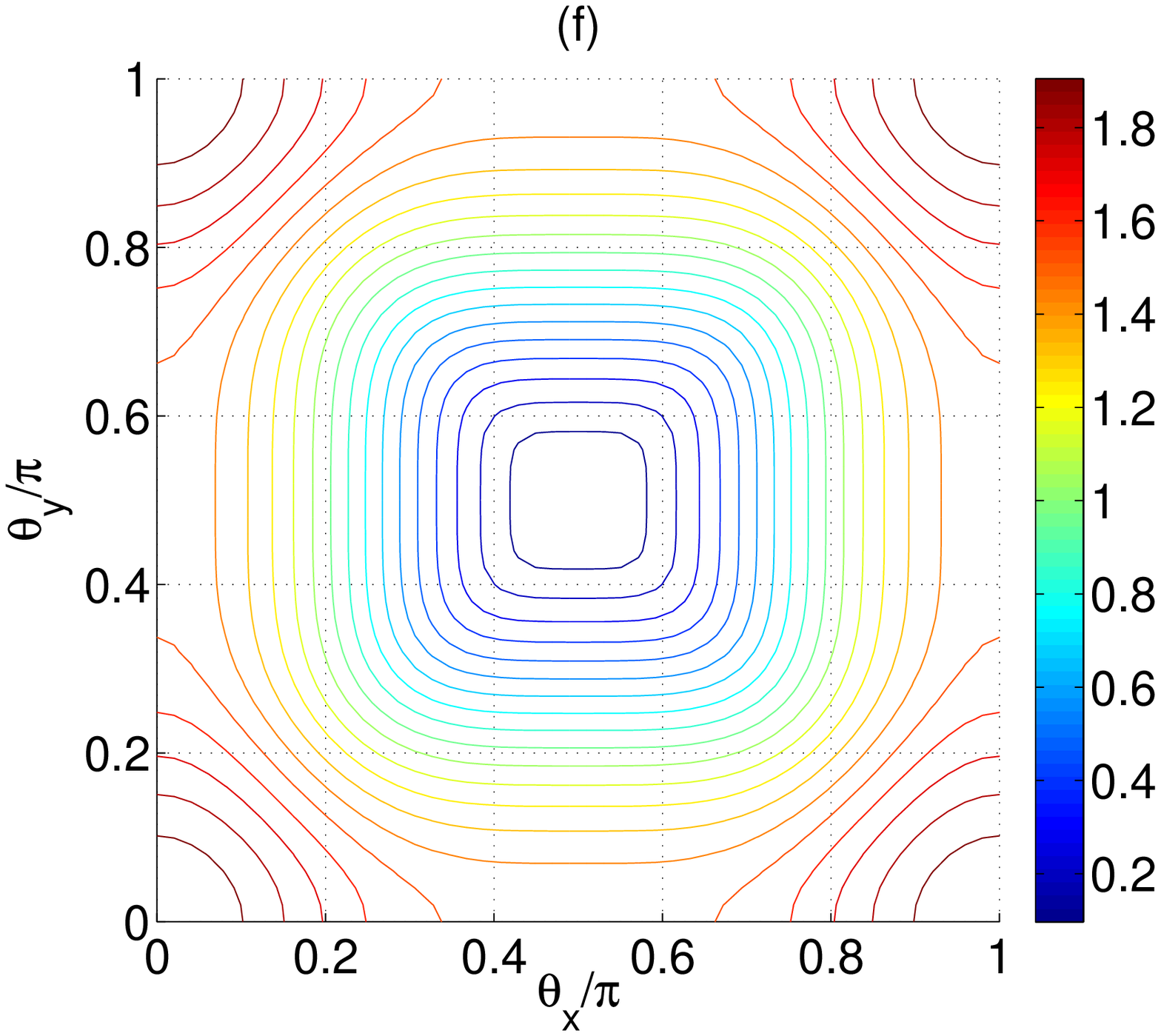}\\
\caption[Fig5]{(Color online) Eigenenergies for a $2\times 2$ 
system with spin-orbit coupling as a function of $\theta_x$ and
$\theta_y$. The inset in (a) shows the solutions at $\theta_y=0$.
We choose $U_{11}=U_{22}=U_{12}=1$ and $t_x=t_y=0.5$. All states
are degenerate at $\theta_x=\theta_y=\pi/2$.
}
\label{fig:SO}
\end{figure}

In a two-dimensional system situation is more complex.
In Fig.~\ref{fig:SO} we show the eigenenergies of the spin-orbit coupled
plaquette as a function of $\theta_x$ and $\theta_y$. 
As SO coupling is turned on the initial three-fold degeneracy
of the first excited state is broken and all states are non-degenerate.
However, remarkably as one approaches $\theta_x=\theta_y=\pi/2$ all states become degenerate.
(For non-interacting system one has at this point three-fold degenerate states
with energies $E_\pm=\pm\sqrt{t_x^2+t_y^2}$.) This suggests that close to this point
the perturbative approach we have used might be breaking down and more accurate
theory might be required to properly resolve the eigenstates. Also, this high 
dimensionality of the ground state manifold might have interesting consequences
for the dynamics as well as on how the system responds to experimental probes.

\section{Conclusions}
\label{sec:conclusions}
In this article we have explored the physics of small insulating plaquettes
under the influence of gauge potentials and spin-orbit couplings.
Computations were done for bosonic systems,
but in the simplest cases two-component fermionic problem amounts to sign change in the eigenenergies.
For example, the highest excited state for bosons can become the ground state for fermions.
At the formal level our approach does not distinguish between bosons and fermions since
the only inputs into the solver are the various onsite interactions strengths and 
tunneling coefficients. For this reason the approach used here is also easy to apply to the study of
fermionic multi-flavor systems~\cite{Krauser2012a} or Bose-Fermi mixtures.
Since atoms could also be prepared on the excited bands of the lattice~\cite{Wirth2011a},
it would be of interest to also explore the Mott insulating plaquettes
in bipartite systems~\cite{Martikainen2012a} and in excited bands.

\begin{acknowledgments}
This work was supported by the Academy of Finland through
its Centres of Excellence Programme (projects No. 251748, No. 263347, No. 135000 and No. 141039)
and the author acknowledges financial support from 
the Academy of Finland (Project 135646).
\end{acknowledgments}

%\bibliographystyle{apsrev}
%\bibliography{./bibli}

\begin{thebibliography}{26}
\expandafter\ifx\csname natexlab\endcsname\relax\def\natexlab#1{#1}\fi
\expandafter\ifx\csname bibnamefont\endcsname\relax
  \def\bibnamefont#1{#1}\fi
\expandafter\ifx\csname bibfnamefont\endcsname\relax
  \def\bibfnamefont#1{#1}\fi
\expandafter\ifx\csname citenamefont\endcsname\relax
  \def\citenamefont#1{#1}\fi
\expandafter\ifx\csname url\endcsname\relax
  \def\url#1{\texttt{#1}}\fi
\expandafter\ifx\csname urlprefix\endcsname\relax\def\urlprefix{URL }\fi
\providecommand{\bibinfo}[2]{#2}
\providecommand{\eprint}[2][]{\url{#2}}

\bibitem[{\citenamefont{Dalfovo et~al.}(1999)\citenamefont{Dalfovo, Giorgini,
  Pitaevskii, and Stringari}}]{Dalfovo1999a}
\bibinfo{author}{\bibfnamefont{F.}~\bibnamefont{Dalfovo}},
  \bibinfo{author}{\bibfnamefont{S.}~\bibnamefont{Giorgini}},
  \bibinfo{author}{\bibfnamefont{L.~P.} \bibnamefont{Pitaevskii}},
  \bibnamefont{and}
  \bibinfo{author}{\bibfnamefont{S.}~\bibnamefont{Stringari}},
  \bibinfo{journal}{Rev. Mod. Phys.} \textbf{\bibinfo{volume}{71}},
  \bibinfo{pages}{463} (\bibinfo{year}{1999}).

\bibitem[{\citenamefont{Greiner et~al.}(2002)\citenamefont{Greiner, Mandel,
  Esslinger, {H\"{a}nsch}, and Bloch}}]{Greiner2002a}
\bibinfo{author}{\bibfnamefont{M.}~\bibnamefont{Greiner}},
  \bibinfo{author}{\bibfnamefont{O.}~\bibnamefont{Mandel}},
  \bibinfo{author}{\bibfnamefont{T.}~\bibnamefont{Esslinger}},
  \bibinfo{author}{\bibfnamefont{T.~W.} \bibnamefont{{H\"{a}nsch}}},
  \bibnamefont{and} \bibinfo{author}{\bibfnamefont{I.}~\bibnamefont{Bloch}},
  \bibinfo{journal}{Nature} \textbf{\bibinfo{volume}{415}}, \bibinfo{pages}{39}
  (\bibinfo{year}{2002}).

\bibitem[{\citenamefont{Sachdev}(2008)}]{Sachdev2008a}
\bibinfo{author}{\bibfnamefont{S.}~\bibnamefont{Sachdev}},
  \bibinfo{journal}{Nature Physics} \textbf{\bibinfo{volume}{4}},
  \bibinfo{pages}{173} (\bibinfo{year}{2008}).

\bibitem[{\citenamefont{Kuklov and Svistunov}(2003)}]{Kuklov2003a}
\bibinfo{author}{\bibfnamefont{A.~B.} \bibnamefont{Kuklov}} \bibnamefont{and}
  \bibinfo{author}{\bibfnamefont{B.~V.} \bibnamefont{Svistunov}},
  \bibinfo{journal}{Phys. Rev. Lett.} \textbf{\bibinfo{volume}{90}},
  \bibinfo{pages}{100401} (\bibinfo{year}{2003}).

\bibitem[{\citenamefont{Altman et~al.}(2003)\citenamefont{Altman, Hofstetter,
  Demler, and Lukin}}]{Altman2003a}
\bibinfo{author}{\bibfnamefont{E.}~\bibnamefont{Altman}},
  \bibinfo{author}{\bibfnamefont{W.}~\bibnamefont{Hofstetter}},
  \bibinfo{author}{\bibfnamefont{E.}~\bibnamefont{Demler}}, \bibnamefont{and}
  \bibinfo{author}{\bibfnamefont{M.~D.} \bibnamefont{Lukin}},
  \bibinfo{journal}{New J. Phys.} \textbf{\bibinfo{volume}{5}},
  \bibinfo{pages}{113} (\bibinfo{year}{2003}).

\bibitem[{\citenamefont{Lin et~al.}(2011)\citenamefont{Lin, Jimenez-García,
  and Spielman}}]{Lin2011a}
\bibinfo{author}{\bibfnamefont{Y.-J.} \bibnamefont{Lin}},
  \bibinfo{author}{\bibfnamefont{K.}~\bibnamefont{Jimenez-García}},
  \bibnamefont{and} \bibinfo{author}{\bibfnamefont{I.~B.}
  \bibnamefont{Spielman}}, \bibinfo{journal}{Nature}
  \textbf{\bibinfo{volume}{471}}, \bibinfo{pages}{83} (\bibinfo{year}{2011}).

\bibitem[{\citenamefont{Struck et~al.}(2011)\citenamefont{Struck,
  \"{O}lschl\"{a}ger, Le~Targat, {Soltan-Panahi}, Eckardt, Lewenstein,
  Windpassinger, and Sengstock}}]{Struck2011a}
\bibinfo{author}{\bibfnamefont{J.}~\bibnamefont{Struck}},
  \bibinfo{author}{\bibfnamefont{C.}~\bibnamefont{\"{O}lschl\"{a}ger}},
  \bibinfo{author}{\bibfnamefont{R.}~\bibnamefont{Le~Targat}},
  \bibinfo{author}{\bibfnamefont{P.}~\bibnamefont{{Soltan-Panahi}}},
  \bibinfo{author}{\bibfnamefont{A.}~\bibnamefont{Eckardt}},
  \bibinfo{author}{\bibfnamefont{M.}~\bibnamefont{Lewenstein}},
  \bibinfo{author}{\bibfnamefont{P.}~\bibnamefont{Windpassinger}},
  \bibnamefont{and}
  \bibinfo{author}{\bibfnamefont{K.}~\bibnamefont{Sengstock}},
  \bibinfo{journal}{Science} \textbf{\bibinfo{volume}{333}},
  \bibinfo{pages}{996} (\bibinfo{year}{2011}).

\bibitem[{\citenamefont{Struck et~al.}(2012)\citenamefont{Struck,
  {\"{O}}lschl{\"{a}}ger, Weinberg, Hauke, Simonet, Eckardt, Lewenstein,
  Sengstock, and Windpassinger}}]{Struck2012a}
\bibinfo{author}{\bibfnamefont{J.}~\bibnamefont{Struck}},
  \bibinfo{author}{\bibfnamefont{C.}~\bibnamefont{{\"{O}}lschl{\"{a}}ger}},
  \bibinfo{author}{\bibfnamefont{M.}~\bibnamefont{Weinberg}},
  \bibinfo{author}{\bibfnamefont{P.}~\bibnamefont{Hauke}},
  \bibinfo{author}{\bibfnamefont{J.}~\bibnamefont{Simonet}},
  \bibinfo{author}{\bibfnamefont{A.}~\bibnamefont{Eckardt}},
  \bibinfo{author}{\bibfnamefont{M.}~\bibnamefont{Lewenstein}},
  \bibinfo{author}{\bibfnamefont{K.}~\bibnamefont{Sengstock}},
  \bibnamefont{and}
  \bibinfo{author}{\bibfnamefont{P.}~\bibnamefont{Windpassinger}},
  \bibinfo{journal}{Phys. Rev. Lett.} \textbf{\bibinfo{volume}{108}},
  \bibinfo{pages}{225304} (\bibinfo{year}{2012}).

\bibitem[{\citenamefont{Ruseckas et~al.}(2005)\citenamefont{Ruseckas,
  Juzeliunas, {\"{O}}hberg, and Fleischhauer}}]{Ruseckas2005a}
\bibinfo{author}{\bibfnamefont{J.}~\bibnamefont{Ruseckas}},
  \bibinfo{author}{\bibfnamefont{G.}~\bibnamefont{Juzeliunas}},
  \bibinfo{author}{\bibfnamefont{P.}~\bibnamefont{{\"{O}}hberg}},
  \bibnamefont{and}
  \bibinfo{author}{\bibfnamefont{M.}~\bibnamefont{Fleischhauer}},
  \bibinfo{journal}{Phys. Rev. Lett.} \textbf{\bibinfo{volume}{95}},
  \bibinfo{pages}{010404} (\bibinfo{year}{2005}).

\bibitem[{\citenamefont{Hauke et~al.}(2012)\citenamefont{Hauke, Tielemann,
  Celi, {\"{O}}lschl{\"{a}}ger, Simonet, Struck, Weinberg, Windpassinger,
  Sengstock, Lewenstein et~al.}}]{Hauke2012a}
\bibinfo{author}{\bibfnamefont{P.}~\bibnamefont{Hauke}},
  \bibinfo{author}{\bibfnamefont{O.}~\bibnamefont{Tielemann}},
  \bibinfo{author}{\bibfnamefont{A.}~\bibnamefont{Celi}},
  \bibinfo{author}{\bibfnamefont{C.}~\bibnamefont{{\"{O}}lschl{\"{a}}ger}},
  \bibinfo{author}{\bibfnamefont{J.}~\bibnamefont{Simonet}},
  \bibinfo{author}{\bibfnamefont{J.}~\bibnamefont{Struck}},
  \bibinfo{author}{\bibfnamefont{M.}~\bibnamefont{Weinberg}},
  \bibinfo{author}{\bibfnamefont{P.}~\bibnamefont{Windpassinger}},
  \bibinfo{author}{\bibfnamefont{K.}~\bibnamefont{Sengstock}},
  \bibinfo{author}{\bibfnamefont{M.}~\bibnamefont{Lewenstein}},
  \bibnamefont{et~al.}, \bibinfo{journal}{{arXiv:1205.1398}}
  (\bibinfo{year}{2012}).

\bibitem[{\citenamefont{Dalibard et~al.}(2011)\citenamefont{Dalibard, Gerbier,
  Juzeliunas, and {\"{O}}hberg}}]{Dalibard2011a}
\bibinfo{author}{\bibfnamefont{J.}~\bibnamefont{Dalibard}},
  \bibinfo{author}{\bibfnamefont{F.}~\bibnamefont{Gerbier}},
  \bibinfo{author}{\bibfnamefont{G.}~\bibnamefont{Juzeliunas}},
  \bibnamefont{and}
  \bibinfo{author}{\bibfnamefont{P.}~\bibnamefont{{\"{O}}hberg}},
  \bibinfo{journal}{Rev. Mod. Phys.} \textbf{\bibinfo{volume}{83}},
  \bibinfo{pages}{1523} (\bibinfo{year}{2011}).

\bibitem[{\citenamefont{Nascimb\`{e}ne
  et~al.}(2012)\citenamefont{Nascimb\`{e}ne, Chen, Atala, Aidelsburger,
  Trotzky, Paredes, and Bloch}}]{Nascimbene2012}
\bibinfo{author}{\bibfnamefont{S.}~\bibnamefont{Nascimb\`{e}ne}},
  \bibinfo{author}{\bibfnamefont{Y.}~\bibnamefont{Chen}},
  \bibinfo{author}{\bibfnamefont{M.}~\bibnamefont{Atala}},
  \bibinfo{author}{\bibfnamefont{M.}~\bibnamefont{Aidelsburger}},
  \bibinfo{author}{\bibfnamefont{S.}~\bibnamefont{Trotzky}},
  \bibinfo{author}{\bibfnamefont{B.}~\bibnamefont{Paredes}}, \bibnamefont{and}
  \bibinfo{author}{\bibfnamefont{I.}~\bibnamefont{Bloch}},
  \bibinfo{journal}{{arXiv:1202.6361}}  (\bibinfo{year}{2012}).

\bibitem[{\citenamefont{Cole et~al.}(2012)\citenamefont{Cole, Zhang,
  Paramekanti, and Trivedi}}]{Cole2012a}
\bibinfo{author}{\bibfnamefont{W.~S.} \bibnamefont{Cole}},
  \bibinfo{author}{\bibfnamefont{S.}~\bibnamefont{Zhang}},
  \bibinfo{author}{\bibfnamefont{A.}~\bibnamefont{Paramekanti}},
  \bibnamefont{and} \bibinfo{author}{\bibfnamefont{N.}~\bibnamefont{Trivedi}},
  \bibinfo{journal}{{arXiv:1205.2319}}  (\bibinfo{year}{2012}).

\bibitem[{\citenamefont{Radic et~al.}(2012)\citenamefont{Radic, Di~Ciolo, Sun,
  and Galitski}}]{Radic2012a}
\bibinfo{author}{\bibfnamefont{J.}~\bibnamefont{Radic}},
  \bibinfo{author}{\bibfnamefont{A.}~\bibnamefont{Di~Ciolo}},
  \bibinfo{author}{\bibfnamefont{K.}~\bibnamefont{Sun}}, \bibnamefont{and}
  \bibinfo{author}{\bibfnamefont{V.}~\bibnamefont{Galitski}},
  \bibinfo{journal}{{arXiv:1205.2110}}  (\bibinfo{year}{2012}).

\bibitem[{\citenamefont{Krauser et~al.}(2012)\citenamefont{Krauser, Heinze,
  Fl{\"{a}}schner, G{\"{o}}tze, Becker, and Sengstock}}]{Krauser2012a}
\bibinfo{author}{\bibfnamefont{J.~S.} \bibnamefont{Krauser}},
  \bibinfo{author}{\bibfnamefont{J.}~\bibnamefont{Heinze}},
  \bibinfo{author}{\bibfnamefont{N.}~\bibnamefont{Fl{\"{a}}schner}},
  \bibinfo{author}{\bibfnamefont{S.}~\bibnamefont{G{\"{o}}tze}},
  \bibinfo{author}{\bibfnamefont{C.}~\bibnamefont{Becker}}, \bibnamefont{and}
  \bibinfo{author}{\bibfnamefont{K.}~\bibnamefont{Sengstock}},
  \bibinfo{journal}{{arXiv:1203.0948}}  (\bibinfo{year}{2012}).

\bibitem[{\citenamefont{Best et~al.}(2009)\citenamefont{Best, Will, Schneider,
  Hackerm{\"{u}}ller, van Oosten, Bloch, and L{\"{u}}hmann}}]{Best2009a}
\bibinfo{author}{\bibfnamefont{T.}~\bibnamefont{Best}},
  \bibinfo{author}{\bibfnamefont{S.}~\bibnamefont{Will}},
  \bibinfo{author}{\bibfnamefont{U.}~\bibnamefont{Schneider}},
  \bibinfo{author}{\bibfnamefont{L.}~\bibnamefont{Hackerm{\"{u}}ller}},
  \bibinfo{author}{\bibfnamefont{D.}~\bibnamefont{van Oosten}},
  \bibinfo{author}{\bibfnamefont{I.}~\bibnamefont{Bloch}}, \bibnamefont{and}
  \bibinfo{author}{\bibfnamefont{D.-S.} \bibnamefont{L{\"{u}}hmann}},
  \bibinfo{journal}{Phys. Rev. Lett.} \textbf{\bibinfo{volume}{102}},
  \bibinfo{pages}{030408} (\bibinfo{year}{2009}).

\bibitem[{\citenamefont{Shin et~al.}(2008)\citenamefont{Shin, Schirotzek,
  Schunck, and Ketterle}}]{Shin2008b}
\bibinfo{author}{\bibfnamefont{Y.-I.} \bibnamefont{Shin}},
  \bibinfo{author}{\bibfnamefont{A.}~\bibnamefont{Schirotzek}},
  \bibinfo{author}{\bibfnamefont{C.~H.} \bibnamefont{Schunck}},
  \bibnamefont{and} \bibinfo{author}{\bibfnamefont{W.}~\bibnamefont{Ketterle}},
  \bibinfo{journal}{Phys. Rev. Lett.} \textbf{\bibinfo{volume}{101}},
  \bibinfo{pages}{070404} (\bibinfo{year}{2008}).

\bibitem[{com()}]{comment2012a}
\bibinfo{note}{In spinorial systems or in higher bands there can be collisions
  that change the atom number of a component. In these cases simple Fock states
  are not the eigenstates of the system with more than one atom per site and
  the intermediate states created by the tunneling operator should be expanded
  in terms of the eigenstates of the interaction Hamiltonian. However, in this
  paper this distinction does not matter.}

\bibitem[{\citenamefont{Martikainen and Larson}(2012)}]{Martikainen2012a}
\bibinfo{author}{\bibfnamefont{J.-P.} \bibnamefont{Martikainen}}
  \bibnamefont{and} \bibinfo{author}{\bibfnamefont{J.}~\bibnamefont{Larson}},
  \bibinfo{journal}{{arXiv:1203.4402}}  (\bibinfo{year}{2012}).

\bibitem[{\citenamefont{Eckardt et~al.}(2010)\citenamefont{Eckardt, Hauke,
  {Soltan-Panahi}, Becker, Sengstock, and Lewenstein}}]{Eckardt2010a}
\bibinfo{author}{\bibfnamefont{A.}~\bibnamefont{Eckardt}},
  \bibinfo{author}{\bibfnamefont{P.}~\bibnamefont{Hauke}},
  \bibinfo{author}{\bibfnamefont{P.}~\bibnamefont{{Soltan-Panahi}}},
  \bibinfo{author}{\bibfnamefont{C.}~\bibnamefont{Becker}},
  \bibinfo{author}{\bibfnamefont{K.}~\bibnamefont{Sengstock}},
  \bibnamefont{and}
  \bibinfo{author}{\bibfnamefont{M.}~\bibnamefont{Lewenstein}},
  \bibinfo{journal}{Europhysics Letters} \textbf{\bibinfo{volume}{89}},
  \bibinfo{pages}{10010} (\bibinfo{year}{2010}).

\bibitem[{\citenamefont{Gerbier and Dalibard}(2010)}]{Gerbier2010a}
\bibinfo{author}{\bibfnamefont{F.}~\bibnamefont{Gerbier}} \bibnamefont{and}
  \bibinfo{author}{\bibfnamefont{J.}~\bibnamefont{Dalibard}},
  \bibinfo{journal}{New Journal of Physics} \textbf{\bibinfo{volume}{12}},
  \bibinfo{pages}{033007} (\bibinfo{year}{2010}).

\bibitem[{\citenamefont{Sacha et~al.}(2012)\citenamefont{Sacha,
  Targo{\'{n}}ska, and Zakrzewski}}]{Sacha2012a}
\bibinfo{author}{\bibfnamefont{K.}~\bibnamefont{Sacha}},
  \bibinfo{author}{\bibfnamefont{K.}~\bibnamefont{Targo{\'{n}}ska}},
  \bibnamefont{and}
  \bibinfo{author}{\bibfnamefont{J.}~\bibnamefont{Zakrzewski}},
  \bibinfo{journal}{Phys. Rev. A} \textbf{\bibinfo{volume}{85}},
  \bibinfo{pages}{053613} (\bibinfo{year}{2012}).

\bibitem[{\citenamefont{Lignier et~al.}(2007)\citenamefont{Lignier, Sias,
  Ciampini, Singh, Zenesini, Morsch, and Arimondo}}]{Lignier2007a}
\bibinfo{author}{\bibfnamefont{H.}~\bibnamefont{Lignier}},
  \bibinfo{author}{\bibfnamefont{C.}~\bibnamefont{Sias}},
  \bibinfo{author}{\bibfnamefont{D.}~\bibnamefont{Ciampini}},
  \bibinfo{author}{\bibfnamefont{Y.}~\bibnamefont{Singh}},
  \bibinfo{author}{\bibfnamefont{A.}~\bibnamefont{Zenesini}},
  \bibinfo{author}{\bibfnamefont{O.}~\bibnamefont{Morsch}}, \bibnamefont{and}
  \bibinfo{author}{\bibfnamefont{E.}~\bibnamefont{Arimondo}},
  \bibinfo{journal}{Phys. Rev. Lett.} \textbf{\bibinfo{volume}{99}},
  \bibinfo{pages}{220403} (\bibinfo{year}{2007}).

\bibitem[{\citenamefont{Chen et~al.}(2011)\citenamefont{Chen, Nascimb\`{e}ne,
  Aidelsburger, Atala, Trotzky, and Bloch}}]{Chen2011a}
\bibinfo{author}{\bibfnamefont{Y.-A.} \bibnamefont{Chen}},
  \bibinfo{author}{\bibfnamefont{S.}~\bibnamefont{Nascimb\`{e}ne}},
  \bibinfo{author}{\bibfnamefont{M.}~\bibnamefont{Aidelsburger}},
  \bibinfo{author}{\bibfnamefont{M.}~\bibnamefont{Atala}},
  \bibinfo{author}{\bibfnamefont{S.}~\bibnamefont{Trotzky}}, \bibnamefont{and}
  \bibinfo{author}{\bibfnamefont{I.}~\bibnamefont{Bloch}},
  \bibinfo{journal}{Phys. Rev. Lett.} \textbf{\bibinfo{volume}{107}},
  \bibinfo{pages}{210405} (\bibinfo{year}{2011}).

\bibitem[{\citenamefont{Ma et~al.}(2011)\citenamefont{Ma, Tai, Preiss, Bakr,
  Simon, and Greiner}}]{Ma2011a}
\bibinfo{author}{\bibfnamefont{R.}~\bibnamefont{Ma}},
  \bibinfo{author}{\bibfnamefont{M.~E.} \bibnamefont{Tai}},
  \bibinfo{author}{\bibfnamefont{P.~M.} \bibnamefont{Preiss}},
  \bibinfo{author}{\bibfnamefont{W.~S.} \bibnamefont{Bakr}},
  \bibinfo{author}{\bibfnamefont{J.}~\bibnamefont{Simon}}, \bibnamefont{and}
  \bibinfo{author}{\bibfnamefont{M.}~\bibnamefont{Greiner}},
  \bibinfo{journal}{Phys. Rev. Lett.} \textbf{\bibinfo{volume}{107}},
  \bibinfo{pages}{095301} (\bibinfo{year}{2011}).

\bibitem[{\citenamefont{Wirth et~al.}(2011)\citenamefont{Wirth, \"Olschl\"ager,
  and Hemmerich}}]{Wirth2011a}
\bibinfo{author}{\bibfnamefont{G.}~\bibnamefont{Wirth}},
  \bibinfo{author}{\bibfnamefont{M.}~\bibnamefont{\"Olschl\"ager}},
  \bibnamefont{and}
  \bibinfo{author}{\bibfnamefont{A.}~\bibnamefont{Hemmerich}},
  \bibinfo{journal}{Nature Physics} \textbf{\bibinfo{volume}{7}},
  \bibinfo{pages}{147} (\bibinfo{year}{2011}).

\end{thebibliography}

\end{document}